\documentclass[prd,a4paper,preprint,showpacs,byrevtex,onecolumn]{revtex4}
\usepackage{graphicx}
\usepackage{amsmath}
\usepackage{array}
\usepackage{bm}
\usepackage{textcomp}

\begin{document}

\title{Effective potential between two gluons from the scalar glueball}

\author{Fabien \surname{Buisseret}}
\thanks{FNRS Research Fellow}
\email[E-mail: ]{fabien.buisseret@umh.ac.be}
\author{Claude \surname{Semay}}
\thanks{FNRS Research Associate}
\email[E-mail: ]{claude.semay@umh.ac.be} 
\affiliation{Groupe de Physique Nucl\'{e}aire Th\'{e}orique, Universit\'{e} de Mons-Hainaut, Acad\'{e}mie universitaire Wallonie-Bruxelles, Place du Parc 20, BE-7000 Mons, Belgium.}

\date{\today}

\begin{abstract}
Starting from the $0^{++}$ glueball mass and wave function computed from
lattice QCD, we compute the local potential between two constituent
gluons. Since the properties of constituent gluons are still a matter of research, we allow for them to be either massless, or massive with a mass around $0.7$~GeV. Both pictures are actually used in the literature. When the gluons are massless, the corresponding local potential is shown to be compatible with a Cornell form, that is a linear
confinement plus a short-range Coulomb part, with standard values for the
flux tube energy density and for the strong coupling constant. When the gluons are massive, the confining potential is a saturating one, commonly used to simulate string-breaking effects. These results fill a gap between lattice QCD and phenomenological models: The picture of the scalar glueball as a bound state of two constituent gluons interacting via a phenomenological potential is shown to emerge from pure gauge lattice QCD computations. Moreover, we show that the allowed potential shape is constrained by the mass of the constituent gluons.
\end{abstract}

\pacs{12.39.Mk, 12.39.Ki, 12.39.Pn}

\maketitle

\section{Introduction}
The study of glueballs currently deserves much interest from a
theoretical point of view, either within the framework of lattice QCD or effective models. Their experimental detection is also an active field of research--for recent reviews, see Refs.~\cite{glurev}. Among the various effective approaches which have been proposed (bag model \cite{bagglu}, QCD in Coulomb gauge \cite{scz}, \dots), many studies have been devoted to potential models of glueballs. In this framework, glueballs are seen as bound states of two or
more constituent gluons interacting via a phenomenological potential. Early works
on this subject are quoted in Refs.~\cite{first,glue2}, and more recent
studies can be found in Refs.~\cite{glue2,simo2,buisnew,brau,glue2_2}.

Actually, the relevance of using a potential model to describe a glueball is still controversial. Assuming that a potential model can be used, two basic questions then appear: What is the mass of a constituent gluon, and what is the potential? The answer to these questions differ from one approach to another. Let us begin by the problem of the mass. On the one hand, it is argued in some works that a gluon is a massless particle, which acquires a dynamical mass given by $\mu=\left\langle \sqrt{\bm p^2}\right\rangle$ because of the confining interaction. Relativistic, spin-dependent, corrections, are then developed in powers of $1/\mu^2$ \cite{first,simo2,buisnew}. In this picture, the constituent gluon is \textit{a posteriori} massive, because it is confined into a glueball, and this mass is state-dependent. It is worth mentioning that, more generally, both quarks and gluons can acquire a constituent mass from renormalization theory. This constituent mass runs with the momentum: One can look in particular at the Coulomb gauge approach of Ref.~\cite{llan1}, where it is shown that massless gluons acquire a running mass which is about $0.7$ GeV at zero momentum. The same approach can also be applied to compute the constituent mass of light quarks~\cite{llan2}. On the other hand, it is often assumed in different studies that a constituent gluon has to be \textit{a priori} considered as massive \cite{glue2,hou}. The idea underlying these last approaches is roughly that the nonperturbative effects of QCD causes a mass term to appear in the gluon propagator. Consequently, the gluons should be seen as massive particles, with a fixed mass which is typically assumed to be around $0.5\pm0.2$~GeV \cite{glue2,aguil}. The relativistic corrections are then expanded as usual in powers of $1/m^2_g$. Interestingly, the typical value of $m_g$ in the second approach is compatible with the dynamical mass $\mu$ for the ground state in the first approach, and with the constituent gluon mass at zero momentum of the Coulomb gauge model of Ref.~\cite{llan1}.     

We turn now our attention to the potentials appearing in the various existing models of QCD. In a two-body system, the best-known phenomenological potential is the Cornell one, which is roughly of the form $ar-\kappa/r$, $r$ being the separation between the confined particles. It is worth mentioning that $ar$ is the energy of a straight string of energy density $a$, also called the flux tube, linking the quark to the antiquark and encoding the confining interaction. The Coulomb part is the lowest order contribution of the one gluon exchange processes. The Cornell potential arises from QCD
in the case of a quark-antiquark bound state, as it can be shown by the Wilson loop
technique \cite{loop}. Lattice QCD computations of the energy between a
static quark-antiquark pair also support this potential
\cite[p.~42]{lat0}. Furthermore, background perturbation theory tells that the
potential between two massless constituent gluons should also be of the Cornell form
\cite{simo}. Bound states of gluons with the Cornell potential have been investigated for example in Refs.~\cite{first,simo2,buisnew,brau}. But, as a linearly rising potential neglects string breaking effects, which have been observed in lattice QCD \cite{break} between static quarks, another confining potential is also often used, that is a saturating one of the form $2m_g(1-{\rm e}^{-r/r_c})$. As for the Cornell one, models built on such a potential have been applied to usual hadrons \cite{gonza}, but also to glueballs \cite{glue2,hou}. Let us note that the short-range part corresponding to massive constituent gluons is not a Coulomb term, but is proportional to the Yukawa potential ${\rm e}^{-m_g r}/r$.  

>From this discussion, we can conclude that the best way of dealing with constituent gluons is still controversial. Consequently, it is of particular interest to try to obtain relevant informations from more fundamental approaches such as lattice QCD. If the mass spectrum of pure gauge QCD--the glueball spectrum--is now accurately computed in lattice QCD \cite{lattice1}, the potential energy between two constituent gluons has been much less studied than the quark-antiquark one. Up to know, the only method to obtain this energy with lattice QCD is to compute the
energy between two static sources, these sources being in the adjoint
representation of SU(3) \cite[p.~69]{lat0}. The Cornell shape
is then favored, as in the quark-antiquark case. However, nowadays, both the masses and wave
functions of glueballs can be computed by lattice calculations
\cite{lat,gluwf}. We propose in this paper a new method for
extracting the potential between two constituent gluons from these lattice QCD data.
Such a method has the conceptual advantage of dealing with ``physical"
glueballs rather than with somewhat artificial static sources. It is a
direct application of the Lagrange mesh procedure that we presented in
Ref.~\cite{lag1}.

Our paper is organized as follows. In sec.~\ref{latti}, we recall the main lattice QCD results concerning the lightest scalar glueball. Then, we describe the method to compute the effective gluon-gluon potential in sec.~\ref{compu}, and we comment our results in sec.~\ref{sec:res}. We finally draw some conclusions in sec.~\ref{conclu}.

\section{Results from lattice QCD}\label{latti}

An SU(3) lattice calculation in glueball spectroscopy shows that the
lightest glueball is a scalar particle, whose quantum numbers are
$J^{PC}=0^{++}$, and whose mass is given by $1.710\pm 0.130$~GeV
\cite{lattice1}. Theoretical arguments also support this point \cite{0pp}. The SU(2) wave function of this scalar glueball has
been first computed in Ref.~\cite{lat}, and its mass was
found to be around
$1.2$~GeV, which is lower than the currently accepted SU(3) value. More
recently, the SU(3) scalar glueball wave function has been computed \cite{gluwf}. In this last work, it is found that
\begin{equation}\label{p1}
m_{0^{++}}=1.680\pm0.046\ {\rm GeV},
\end{equation}
in agreement with the result of Ref.~\cite{lattice1}.

\begin{center}
\begin{figure}[ht]
  \resizebox{0.6\textwidth}{!}{%
  \includegraphics{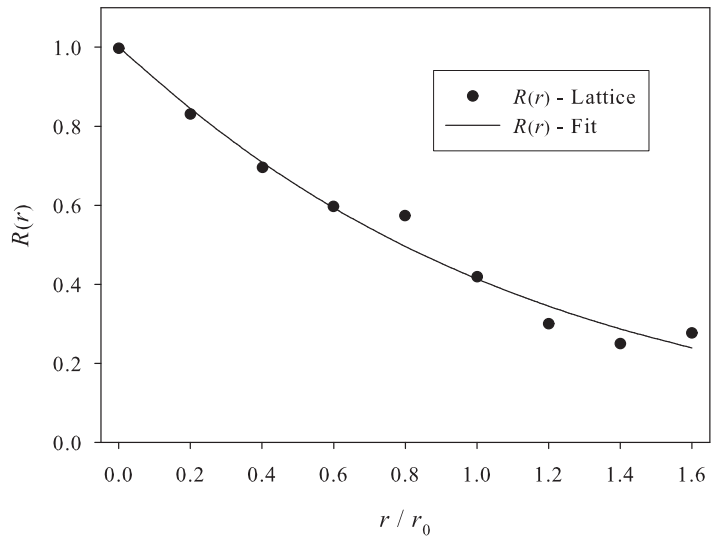}
  }
\caption{Plot of the $0^{++}$ radial wave function, taken from
Ref.~\cite{gluwf} (circles), with $R(0)=1$ and
$r_0=1.472$~GeV$^{-1}$. The fitted radial wave function~(\ref{fit}) is
also plotted (solid line).}
\label{Fig1}
\end{figure}
\end{center}

The $0^{++}$ radial wave function which is computed in Ref.~\cite{gluwf}
seems thus to be a reliable result. As only a few points of
this wave function are available, it is more convenient for latter
calculations to fit them by the trial function
\begin{equation}\label{fit}
  R(r)=\exp\left[-A\left(\frac{r}{r_0}\right)^B\right].
\end{equation}
The size parameter $r_0=0.29$~fm $=1.472$~GeV$^{-1}$ is interpreted as the glueball radius in Ref.~\cite{gluwf}. Its introduction allows to deal with dimensionless fit parameters $A$ and $B$. A fit with
the Levenberg-Marquardt algorithm gives
\begin{eqnarray}\label{p2}
  A=0.883\pm0.045,\ \
  B=1.028\pm0.132,
\end{eqnarray}
with a satisfactory agreement since the coefficient of determination is equal to $0.958$, close to the optimal value of $1$. The result is plotted in Fig.~\ref{Fig1}. Let
us note that $R(r)$ is normalized in such a way that $R(0)=1$.

Consequently, the lattice QCD calculations of Ref.~\cite{gluwf} provide us not only
with the $0^{++}$ glueball mass $m_{0^{++}}$ (\ref{p1}), but also with
a radial wave function, denoted as $R(r)$ and given by
Eqs.~(\ref{fit}) and (\ref{p2}). 

\section{Effective potential}\label{compu}
\subsection{Main equation}

In this work, we want to find a Hamiltonian $H$ such that
\begin{equation}\label{eigen}
  H\,R(r)=m_{0^{++}}\,R(r),
\end{equation}
where $R(r)$ and $m_{0^{++}}$ come from lattice QCD. This eigenequation is the main equation to be solved in the following. To our knowledge, this problem has never been addressed before. It is of particular interest in order to check the relevance of the approaches with constituent gluons. 

In terms of potential models, a glueball can be a bound state of two or more
constituent gluons coupled in a color singlet. We point out that all the potential models agrees with the fact that the lowest $0^{++}$ should be a  mainly two-gluon bound state \cite{first,glue2,simo2,buisnew,brau}. Moreover, the quasiparticle picture of Ref.~\cite{scz}, based on the QCD Hamiltonian in Coulomb gauge, also favors a bound state of two constituent gluons for the lowest $0^{++}$. We will thus assume for $H$ a two-body standard form, i.e. 
\begin{subequations}\label{hamdef}
\begin{equation}
H=T(\bm p^{\,2})+V(r),
\end{equation}
with a local potential $V(r)$ and a semirelativistic kinetic term given by
\begin{equation}\label{Tdef}
T(\bm p^{\,2})=2\ \sqrt{\bm p^{\, 2}+m^2_g}.
\end{equation}
\end{subequations}
Once the mass of the gluon, $m_g$, as well as the radial and orbital quantum numbers $n$ and $\ell$ are specified ($n=\ell=0$ for the lightest scalar glueball), the only unknown quantity in Eq.~(\ref{eigen}) is the local central
potential $V(r)$, that we will compute in the following.

\subsection{Interpretation of the wave function} 

It is worth underlining an important assumption that we make in this work: Motivated by the standard representation of the $0^{++}$ glueball as a two-gluon bound state, we identify the lattice wave function $R(r)$, computed from plaquette operators, with the gluon-gluon component of the scalar glueball wave function. This hypothesis can be intuitively understood as follows. 

A completely relativistic wave equation for a two-body system is provided by the well-known Bethe-Salpeter equation \cite{bethe}. In this formalism, the relativistic wave function $\chi({\rm r})$ (also known as the Bethe-Salpeter wave function) for a bound state $\left|\Omega\right\rangle$ of two particles $\alpha$ and $\beta$ is defined as~\cite{wick}
\begin{equation}\label{bswf1}
	\chi({\rm r})={\rm e}^{-i{\rm P}{\rm R}}\left\langle 0\right| a_\alpha({\rm x}_\alpha)\, a_\beta({\rm x}_\alpha+{\rm r})\left|\Omega\right\rangle.
\end{equation}
In this last equation, $\left|0\right\rangle$ is the vacuum state and $a_\gamma$ is the creation operator for a particle $\gamma$. Moreover, ${\rm x}_\gamma$ is the spacetime coordinate of particle $\gamma$, ${\rm r}={\rm x}_\beta-{\rm x}_\alpha$, and ${\rm P}$ and ${\rm R}$ are the $4$-momentum and the spacetime coordinate of the center of mass of the system respectively. The relativistic wave function~(\ref{bswf1}) can easily be linked to a Schr\"{o}dinger-like wave function. Let us indeed work in the rest frame of the bound state $\left|\Omega\right\rangle$. Then, ${\rm P}=(M_\Omega,\bm 0)$. In the usual nonretarded approximation ${\rm x}^0_a={\rm x}^0_b$, one has
\begin{equation}\label{bswf2}
	\chi({\bm r})={\rm e}^{-i M_\Omega\, t}\left\langle 0\right| a_\alpha({\bm x}_\alpha)\, a_\beta({\bm x}_\alpha+{\bm r})\left|\Omega\right\rangle.
\end{equation}
Since the time coordinate $t$ only appears as a global phase, only the spatial separation $\bm r$, computed in the rest frame of the configuration, is now relevant. 

Formula~(\ref{bswf2}) can then straightforwardly be applied to the gluon-gluon component of a glueball state $\left|G\right\rangle$. Indeed, if $A_\mu$ denotes the creation operator of a gluon, the corresponding Bethe-Salpeter wave function at $t=0$ is given by \cite{lat}
\begin{equation}\label{bswf3}
	\chi(\bm r)=\left\langle0 \right|s^{\mu\nu}\int d\hat{ \bm r}\, Y_{\ell m}(\hat{\bm r})\, A^\dagger_\mu({\bm x})\, A_\nu(\bm x+\bm r)\left|G \right\rangle,
\end{equation}
where the summation on $s^{\mu\nu}$ and the integration on the angular part $\hat {\bm r}$ enforce a particular spin and angular symmetry of the wave function, following the state which is considered. Equation~(\ref{bswf3}) can thus be identified as the gluon-gluon wave function for a particular stationary glueball state, in the rest frame of the system. 

Interestingly, the wave function~(\ref{bswf3}) is precisely what is computed in the lattice QCD studies of Refs.~\cite{lat,gluwf} for the lightest glueball states. The key point to perform such a calculation is the evaluation of the two-gluon operator $A^\dagger_\mu({\bm x})\, A_\nu(\bm x+\bm r)$, which is achieved on the lattice by computing the correlation matrix between two plaquettes at the different points ${\bm x}$ and $\bm x+\bm r$. From this discussion, we can justify the identification of the lattice wave function of Ref.~\cite{gluwf} with the Schr\"{o}dinger-like wave function of a bound state made of two gluons. 

\subsection{Numerical method}

Several methods exist to compute the equivalent local potential from a
given wave function and its corresponding energy (see Ref.~\cite{loc}).
But, these methods are only applicable to the case of nonrelativistic
kinematics. We gave in Ref.~\cite{lag1} a procedure, relying on the
Lagrange mesh method, to make such computations with a semirelativistic
kinematics of the form~(\ref{Tdef}). We recall here the main points of
this method, but refer the reader to Ref.~\cite{lag1} for a detailed
study.

The Lagrange mesh method is a very accurate and simple numerical
procedure to
compute eigenvalues and eigenfunctions of a two-body Schr\"{o}dinger
equation \cite{baye86,vinc93}. It is also applicable to a
semirelativistic kinetic operator, i.e. the spinless Salpeter equation
\cite{sem01}. In the case of radial equations, a Lagrange mesh is
formed of $N$ mesh points $x_{i}$ which are the zeros of
the Laguerre polynomial $L_{N}(x)$ of degree $N$ \cite{baye86}. The
Lagrange basis is then given by a set of $N$ regularized Lagrange
functions,
\begin{equation}
\label{flag}
f_{i}(x)=(-1)^{i}x^{-1/2}_{i}\, x(x-x_{i})^{-1}L_{N}(x)\, e^{-x/2},
\end{equation}
satisfying the condition $f_{j}(x_{i})=\lambda^{-1/2}_{i}\delta_{ij}$
\cite{baye86} and $f_i(0)=0$.
The weights $\lambda_{i}$ are linked to the mesh points $x_{i}$ through
a Gauss quadrature formula
which is used to compute all the integrals over the interval
$[0,\infty[$, that is
\begin{equation}
\label{gauss}
\int^{\infty}_{0} g(x)\, dx  \approx\sum^{N}_{k=1}\lambda_{k}\,
g(x_{k}).
\end{equation}
The regularized wavefunction, given by $u(r)=r\, R(r)$, is then
developed in the Lagrange basis. The semirelativistic kinetic matrix
elements $T_{ij}$ for the operator~(\ref{Tdef}) can be accurately
computed in this basis \cite{sem01}.

We showed in Ref.~\cite{lag1} that, starting from a given regularized wave function
$u(r)=r\, R(r)$, obtained in our case from Eq.~(\ref{fit}), and its corresponding
energy $m_{0^{++}}$, the equivalent local potential is accurately given
at the mesh points by
\begin{equation}\label{effpot1}
  V(hx_i)=m_{0^{++}}-\frac{1}{\sqrt{\lambda_i}\ u(hx_i)}\sum^N_{j=1}T_{
  ij}\sqrt{\lambda_j}\ u(hx_j).
\end{equation}
In the above equation, $h$ is a scale parameter chosen to adjust the
size of the mesh to the domain of physical interest.
The angular orbital momentum $\ell$ has to be a priori specified in
Eq.~(\ref{effpot1}) since the matrix elements $T_{ij}$ depend on $\ell$
\cite{lag1}.
As we deal with the lightest glueball, we assume $\ell=0$ and $S=0$ in
order to obtain a $0^{++}$ state. Equation~(\ref{effpot1}) requires the
knowledge of the wave function at the mesh points $hx_i$. So, it is more
efficient to work with the trial wave function~(\ref{fit}) than with
interpolated points between the few available lattice data.

The first numerical parameter is the number of mesh points, $N$, that we
set equal to $100$ (although $N=30$ already gives a good picture of the
potential \cite{lag1}). The second parameter is the scale parameter $h$.
In Eq.~(\ref{effpot1}), we use a dimensionless variable $x$, with
$r=h\,x$. A relevant value of $h$ is obtained thanks to the relation
$h=r_a/x_N$, where $x_N$ is the last mesh point and $r_a$ is a physical
radius located in the asymptotic tail of the wave function. This radius
has to be a priori estimated, but not with a great
accuracy, since the method is not variational in $h$ \cite{lag1,sem01}.
To determine $h$, we impose that
$R(hx_N)/\max\left[R(r)\right]=\epsilon$, with $R(r)$ given by
Eq.~(\ref{fit}), and $\epsilon$ a small number that we arbitrarily fix
at $\epsilon=10^{-3}$. This way of estimating $h$ has already given good
results \cite{lag1}. As it can be seen from Eq.~(\ref{fit}),
$\max\left[R(r)\right]=1$, and we obtain
\begin{equation}\label{hdef}
  h=\frac{r_0}{x_N}\left[-\frac{\ln \epsilon}{A}\right]^{1/B}.
\end{equation}

Let us note that, strictly speaking, the effective potential $V(r)$ is not unique. Indeed, if $\left|\phi\right\rangle$ is a state such that $\left\langle \phi\right.\left|R\right\rangle=0$ with $\left\langle \bm r\right.\left|R\right\rangle=R(r)$, then the whole class of potentials $U(r)=V(r)+\lambda \left|\phi\right\rangle\left\langle \phi\right|$ satisfies the eigenequation~(\ref{eigen}), $\lambda$ being an arbitrary number. But, these potentials are clearly nonlocal. Up to our knowledge, there is no physical indication that the interaction between two gluons could be non local. We thus compute the unique local potential $V(r)$ for the $0^{++}$ state that we are studying. Consequently, the only possibility is $\lambda=0$, for which $U(r)=V(r)$. 

\section{Results}\label{sec:res}
\subsection{Massless gluons}\label{massless}

\begin{center}
\begin{figure}[ht]
  \resizebox{0.6\textwidth}{!}{%
  \includegraphics{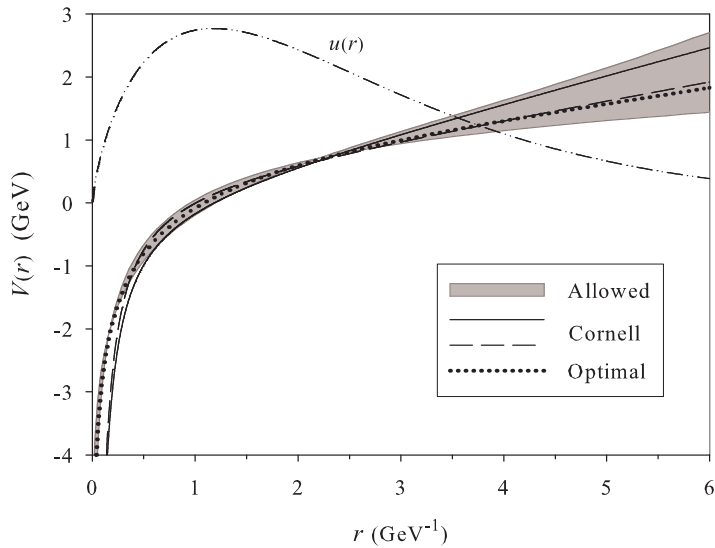}
  }
\caption{Plot of the optimal potential given by Eq.~(\ref{effpot1})
(dotted line) for $m_g=0$. It is computed from the optimal wave function $u(r)$,
plotted
with an arbitrary normalization (dashed-dotted line). The errors on the
glueball mass, $m_{0^{++}}$, and on the wave function parameters $A$ and
$B$ actually allow every potential which is located in the gray area.
These results are compared with two Cornell potentials~(\ref{corn}) for
the standard values $\sigma=0.19$~GeV$^2$ and $\alpha_S=0.20$:
${\cal C}=9/4,\, D=0$ (solid line);
${\cal C}=3/2,\, D=0.30$~GeV (dashed line).}
\label{Fig2}
\end{figure}
\end{center}

We begin by setting $m_g=0$ in the kinetic operator~(\ref{Tdef}). This corresponds to a model in which the gluons are assumed to be massless. With the scale parameter defined by Eq.~(\ref{hdef}), we can apply
formula~(\ref{effpot1}). Results are plotted in Fig.~\ref{Fig2}.

The potential obtained by using the optimal values of $m_{0^{++}}$, $A$, and
$B$ clearly exhibits a confining long-range part, and a rapidly
decreasing short-range part. The errors on these three parameters (see
Eqs.~(\ref{p1}) and (\ref{p2})) allow the ``true" potential to be
located between two extremal curves. We have checked that the curves
obtained remain stable for different values of $\epsilon$, $N$, and $h$.
These numerical results are compatible with the following Cornell
potential
\begin{equation}\label{corn}
  V_C(r)={\cal C}\,\sigma\,r-\frac{3\alpha_S}{r}+D.
\end{equation}
In this expression, $\alpha_S$ is the strong coupling constant, that we set equal to $0.20$ as in the case of the static quark-antiquark potential \cite[p. 42]{lat0}. The $3$ factor is the color factor coming from the one gluon exchange
between two gluons when the pair is in a color singlet. In the confining
part, $\sigma$ is the fundamental quark-antiquark flux tube energy
density. It is usually assumed to be around $0.19$~GeV$^2$ \cite[p. 9]{lat0}.
The constant ${\cal C}$ indicates the scaling of the energy density
which is different for a gluon-gluon system or a quark-antiquark pair. It
is generally assumed to be given by $9/4$ (Casimir scaling)
\cite{cas},
or by $3/2$ (square root of Casimir scaling) \cite{pcas}. The last
constant, $D$, is used to fit the height of potential~(\ref{corn}) on the numerically computed optimal potential. 

The Cornell potential with ${\cal C}=3/2$ is closer to the optimal
curve than the one with the Casimir scaling. However, the interaction
with ${\cal C}=3/2$ demands that $D=0.30$~GeV. This positive constant is
somewhat surprising. Indeed, one expects that the parameter $D$ encodes
some relativistic corrections to the Cornell potential. Talking of such relativistic corrections, like spin-orbit or spin-spin terms, is meaningful even in the case of massless gluons, since these corrections are expressed in terms of powers of $1/\mu^2$, with a rather large value for $\mu=\left\langle \sqrt{\bm p^{\, 2}}\right\rangle$, the dynamical gluon mass generated by the confinement \cite{glu1}. Indeed, a computation of $\mu$ with the fitted wave function~(\ref{fit}) leads to $\mu= 0.585\pm0.128$~GeV.  Since we consider here
a $\ell=S=0$ state, the only remaining relativistic corrections are retardation
terms and contact (spin-spin) interactions, which are both negative in this
case \cite{first,glu2}. The Casimir scaling seems thus more satisfactory
since it is compatible with a vanishing value of $D$ and it is still
located in the allowed region. 

One can see in Fig.~\ref{Fig2} that the
very short-range part of the numerically computed potential is less
singular than the Coulomb potential. This part is actually very
sensitive to the short-range behavior of the wave function. It has been
shown that the wave function of a true semirelativistic Coulomb problem
diverges in $r=0$ \cite{div}, but such a divergence cannot be computed
in lattice calculations. Anyway, more points should be necessary to
elucidate this short-range behavior. 

Let us note that the ground state
masses of the Hamiltonian~(\ref{hamdef}) with the two fitted
potentials~(\ref{corn}) are respectively $1.675$~GeV and $1.670$~GeV
with ${\cal C}=9/4$ and ${\cal C}=3/2$. These values are contained in the error bars of $m_{0^{++}}$. The corresponding wave function with ${\cal C}=3/2$ is nearly indistinguishable of the optimal wave function~(\ref{fit}), while the one for ${\cal C}=9/4$ decreases more quickly. This is coherent with Fig.~\ref{Fig2}, where we can see that the Cornell potential with the Casimir scaling increases faster than the optimal curve, consequently leading to a faster decreasing wave function. 

\subsection{Massive gluons}

\begin{center}
\begin{figure}[ht]
  \resizebox{0.6\textwidth}{!}{%
  \includegraphics{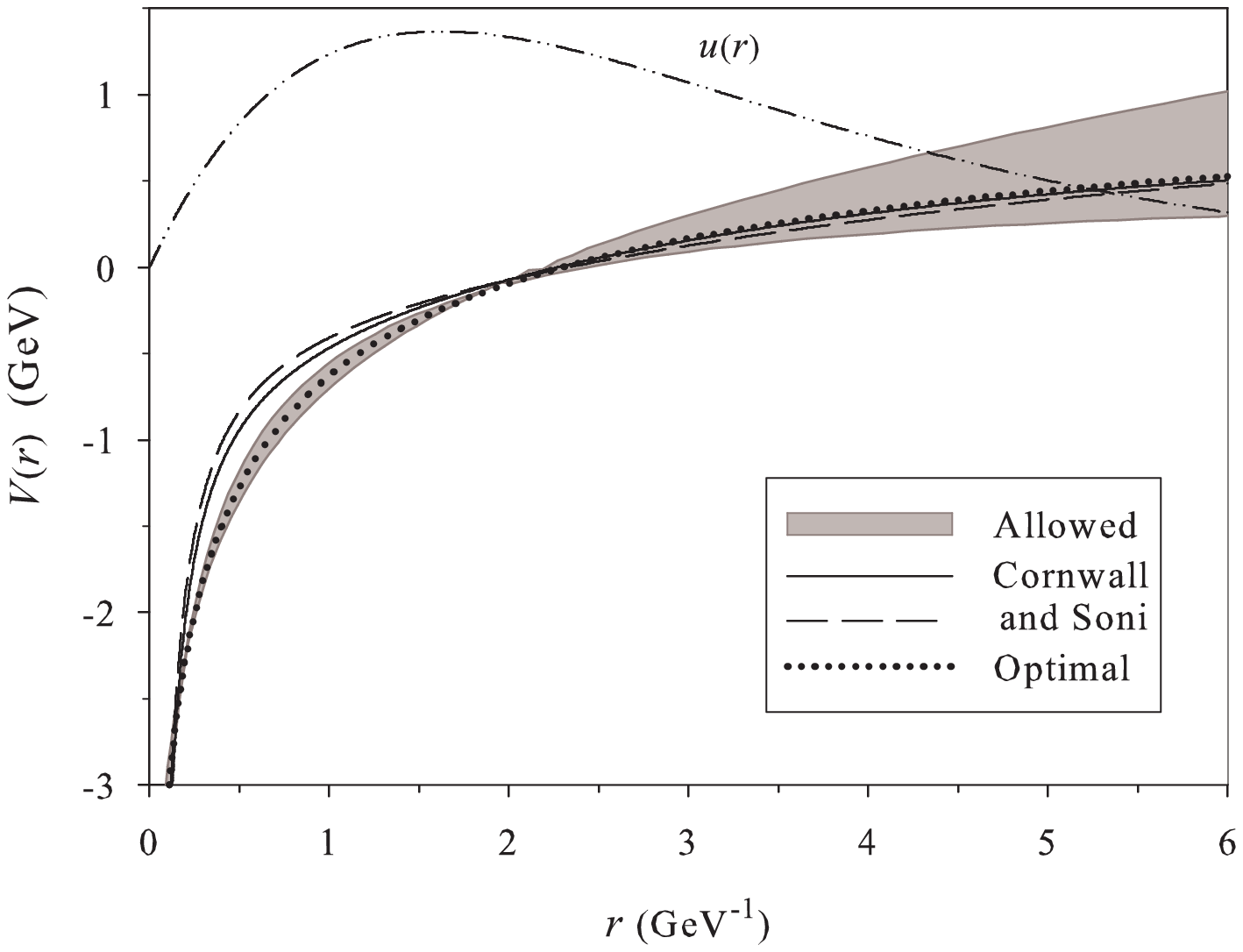}
  }
\caption{Plot of the optimal potential given by Eq.~(\ref{effpot1})
(dotted line) for $m_g=0.7$~GeV. It is computed from the optimal wave function $u(r)$,
plotted
with an arbitrary normalization (dashed-dotted line). The errors on the
glueball mass, $m_{0^{++}}$, and on the wave function parameters $A$ and
$B$ actually allow every potential which is located in the gray area.
These results are compared with two ``Cornwall and Soni potentials" of the form~(\ref{cornw}) with ${\bm S}^2=0$, $r_c=2m_g/{\cal C}\sigma$, $\sigma=0.19$~GeV$^2$, and $\alpha_S=0.67$:
${\cal C}=9/4,\, D=-0.67$~GeV (solid line);
${\cal C}=3/2,\, D=-0.50$~GeV (dashed line).}
\label{Fig3}
\end{figure}
\end{center}

As mentioned in the introduction, it is argued in many works that the constituent gluons should have a fixed nonzero mass, typically around $0.5\pm0.2$~GeV \cite{glue2,aguil}. The effective gluon-gluon potential can also be computed by using our method. In order to clearly see the changes between $m_g=0$ and $m_g>0$, we have computed the effective potential for a rather large but still relevant value of $m_g$, namely $0.7$~GeV. The result is plotted in Fig.~\ref{Fig3}.

A comparison with Fig.~\ref{Fig2} shows that the asymptotic behavior of $V(r)$ drastically changes when $m_g>0$. Instead of a monotonically increasing potential, it now seems to saturate at some energy scale. Such a saturation suggests an interpretation in terms of string breaking effects, which have been measured in lattice QCD in the case of a static quark-antiquark pair. Thus, the optimal potential of Fig.~\ref{Fig3} should no longer be compared to a Cornell form. However, in a pioneering work about glueballs as bound states of massive gluons by Cornwall and Soni \cite{glue2}, it is suggested that the gluon-gluon potential at lowest order reads
\begin{equation}\label{cornw}
	V(r)=2m_g(1-{\rm e}^{-r/r_c})-\alpha_s \left(\frac{1}{2}+\bm S^2\right)\frac{{\rm e}^{-m_gr}}{r}+D,
\end{equation}
where $\bm S$ is the spin of the glueball. This potential clearly saturates at a value $2m_g+D$, then forbidding bound states with a mass greater than $4m_g+D$. For $r\ll r_c$, the confining part reduces to $(2m_g/r_c)\, r$, and a linearly rising confining potential is recovered. By comparison with the Cornell potential~(\ref{corn}), one could expect that 
\begin{equation}
	\frac{2m_g}{r_c}={\cal C} \sigma,
\end{equation}
 which provides us with a definition of $r_c$ in terms the more intuitive parameters $m_g$ and $\sigma$. Some useful remarks can also be made about the short-range part of the potential~(\ref{cornw}), i.e the part which is proportional to $\alpha_s$. Firstly, it involves a Yukawa potential instead of a Coulomb one, because both exchanged gluons and constituent gluons are massive. Moreover, it is spin-dependent even at the lowest order. This is a characteristic feature of effective potential with massive gluons. Other spin-dependent corrections in $1/m^2_g$ also exist, but we do not mention them since they are absorbed in the constant $D$ by definition. Again, this constant is fitted so that the absolute height of the optimal potential is recovered.  
 
It is readily observed in Fig.~\ref{Fig3} that potential~(\ref{cornw}) fits rather well the numerically computed potential, in particular the long-range behavior, for $r> 2$~GeV$^{-1}$. For comparison, we notice that $r_c=3.27$~GeV$^{-1}$ when ${\cal C}=9/4$ and $4.91$~GeV$^{-1}$ when ${\cal C}=3/2$. But, as we argued in the previous section, the short-range part is only poorly known with the current lattice data. The difference between ${\cal C}=9/4$ and $3/2$ is less important than for the Cornell potential even if the Casimir scaling seems closer to the optimal curve when the usual value $\sigma=0.19$~GeV$^2$ is used. It is worth mentioning that we fitted $\alpha_s=0.67$, which is higher than in the case of massless gluons. But, such a value is commonly used in models with massive gluons \cite{glue2,hou}. 

We finally note that the ground state
masses of the Hamiltonian~(\ref{hamdef}) with the two fitted
potentials~(\ref{cornw}) are respectively $1.724$~GeV and $1.721$~GeV
with ${\cal C}=9/4$ and ${\cal C}=3/2$. Again, these values are contained in the error bars of $m_{0^{++}}$. In this case, the corresponding wave functions with both values of ${\cal C}$ are very similar to each other, but with a slightly larger spatial extension than the optimal wave function~(\ref{fit}).
 
\subsection{Mass dependence of the potential} 
 
\begin{center}
\begin{figure}[ht]
  \resizebox{0.6\textwidth}{!}{%
  \includegraphics{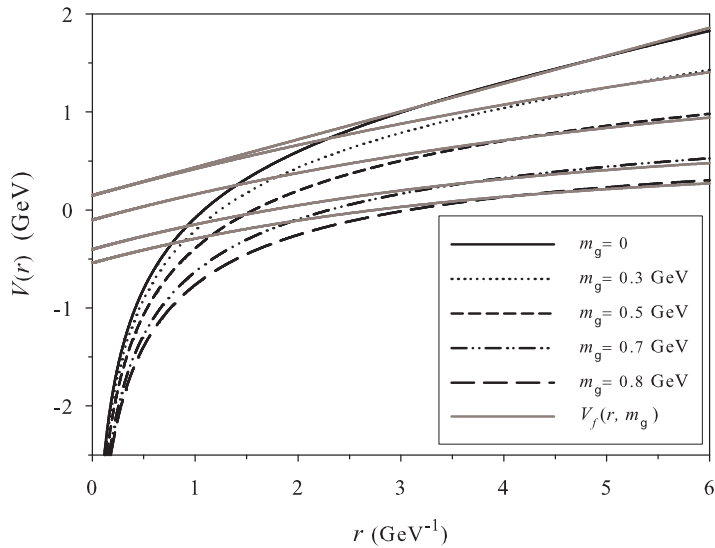}
  }
\caption{Plot of the optimal potential given by Eq.~(\ref{effpot1})
for different values of $m_g$ (black lines). Potentials are computed from the optimal values of the wave function $u(r)$ and of the mass $m_{0^{++}}$. The long-range potential $V_f(r,m_g)$, given by Eq.~(\ref{vfit}), is plotted for the different values of $m_g$ (gray lines) with ${\cal C}=3/2$, $\sigma=0.19$~GeV$^2$, and $\gamma=0.362$. The constant $D$ is fitted so that the absolute height of the corresponding numerically computed potential is well reproduced. }
\label{Fig4}
\end{figure}
\end{center}

A plot of the effective potential for different values of $m_g$ is given in Fig.~\ref{Fig4} in order to see more clearly the evolution of the potential shape with an increasing gluon mass. It is readily observed that there is a transition between a linearly rising regime at zero mass and a saturating one at high gluon mass. 
 
As we already pointed out, the short-range behavior of the computed potentials is rather poorly known because of the current precision of the lattice data. However, interesting informations can be deduced concerning the long-range part of the potential, typically for $r>3$~GeV$^{-1}$. One can find indeed that the different potentials computed in Fig.~\ref{Fig4} can be correctly described at large $r$ by the following form 
\begin{equation}\label{vfit}
	V_f(r,m_g)=\frac{{\cal C}\sigma}{\gamma\, m_g}\left(1-{\rm e}^{-\gamma\, m_g\, r}\right)+D, 
\end{equation}
with $\gamma=0.362\pm0.021$ a fitted parameter. Such a form has the advantage of having a non trivial limit for $m_g=0$, which is simply the linearly rising potential ${\cal C}\sigma r+D$. When $m_g>0$ however, it saturates at the value $({\cal C}\sigma/\gamma \, m_g)+D$. Potential~(\ref{vfit}) has been plotted for the different values of $m_g$ that we considered, with the previously found values of $\sigma$, that is $0.19$~GeV$^2$. We fixed ${\cal C}=3/2$ because we already saw in sec.~\ref{massless} that the optimal numerically computed potential was better reproduced with that value of ${\cal C}$. But, again, we stress that the precision on the potential does not allow to decide whose scaling law is the best one. The constant $D$ have been fitted for each $m_g$ in order for the potential~(\ref{vfit}) to have the right absolute height. 

In the previous section, we showed that the potential obtained with $m_g=0.7$~GeV can be well reproduced by the form~(\ref{cornw}), whose long-range part is a priori inequivalent to the potential~(\ref{vfit}). Actually, both expressions coincide if 
\begin{equation}
	m_g=\sqrt{\frac{{\cal C}\sigma}{2\gamma}}.
\end{equation}
With $\sigma=0.19$~GeV$^2$, this relation states that potentials~(\ref{cornw}) and (\ref{vfit}) are identical for $m_g=0.768\pm0.024$~GeV when ${\cal C}=3/2$, and $m_g=0.627\pm0.020$~GeV for ${\cal C}=9/4$. Remarkably, this corresponds to a gluon mass around $0.7$~GeV, that is the case that we treated previously.

\section{Conclusions}\label{conclu}

In this work, we have shown how to compute the effective potential
between two gluons from the mass and wave function of the $0^{++}$ state
obtained in lattice QCD, reasonably assuming that the lattice wave function mainly gives the gluon-gluon part of the glueball Schr\"{o}dinger-like wave function. This method, which is here used for the
first time, has the advantage of allowing to deal with semirelativistic
kinematics, which is necessary to include massless gluons systems in the discussion.

The relevance of potential models to describe gluons is still a matter of controversy nowadays. In particular, the mass of a constituent gluon is an open problem: Should it be zero or not? Arguments favoring both hypothesis can be found in the literature, so we computed the effective gluon-gluon potential in both cases. When the constituent gluons are massless, the potential we find is compatible with a Cornell one for standard values of the parameters. When the constituent gluons are massive however, the effective potential is merely compatible with a saturating one with a Yukawa-type short-range part. 

These results go beyond the usual computation of the static potential between two static sources in the adjoint representation. In this work indeed, we started from pure gauge results in lattice QCD and we showed that the constituent gluon picture naturally emerges from these lattice data. This could yield an a posteriori justification of the success of potential models in the description of the glueball spectrum. Moreover, we showed that the allowed shape of the potential is constrained by the mass of the constituent gluon (see Eq.~(\ref{vfit})): The linear confinement or the saturating one are only valid for massless or massive gluons respectively. Although the Casimir scaling seems favored, the current precision of the lattice data does not
allow to decide which scaling law is correct. This problem could be
solved by making high precision computations of the wave function. We mention however the recent work of Ref.~\cite{scabicu}, where the behavior of a quark-antiquark-constituent gluon system is investigated on the lattice, with results in agreement with the Casimir scaling hypothesis. 

An important application of our method is that, provided accurate wave
functions for glueballs with a higher total spin are computed (too few
points are currently available for the $2^{++}$ wave
function \cite{gluwf}), spin-dependent terms--vanishing for the scalar glueball--could be studied. This would be a check of the relevance of approaches with massless or massive constituent gluons. Indeed, potentials coming from both models already differ at the lowest order for what concerns spin-dependent terms. We consequently think that a very accurate determination of the wave functions of the lightest $0^{++}$ and $2^{++}$ glueballs in lattice QCD, especially at short range, could serve as a check to determine the most relevant potential approach of glueballs. We expect that, provided these wave functions are known, our method should lead to numerically computed potentials with a common long-range behavior--confinement is indeed not supposed to depend on $J^{PC}$. On the contrary, the short-range behavior should be different, and give important informations about the relativistic corrections. We hope to present such a study in future works.  

\acknowledgments 
The authors thank the FNRS for financial support, and Francis Michel for stimulating discussions.

\end{document}